\documentclass{aa}  
\usepackage{graphicx}
\usepackage{txfonts}
\usepackage[colorlinks]{hyperref}
\usepackage{amsmath}
\usepackage{color}

\newcommand{\kms}{km s$^{-1}{}{}$}
\newcommand{\wm}{W m$^{-2}{}{}$}

\usepackage{array}
\usepackage{tabulary}
\usepackage{rotating}

\begin{document}

         \title{Solar wind energy flux observations in the inner heliosphere: First results from Parker Solar Probe}
         \titlerunning{Solar wind energy flux observations from PSP}


         \author{Mingzhe Liu\inst{1}
                                        \and Karine Issautier\inst{1}
                                        \and Nicole Meyer-Vernet\inst{1}
                                        \and Michel Moncuquet\inst{1}
                                        \and Milan Maksimovic\inst{1}
                                        \and J. S. Halekas\inst{2}
                                        \and J. Huang\inst{3}
                                        \and L. Griton\inst{4}
                                        \and S. Bale\inst{5,6,7,8}
                                        \and J. W. Bonnell\inst{5}
                                        \and A. W. Case\inst{9}
                                        \and K. Goetz\inst{10}
                                        \and P. R. Harvey\inst{5}
                                        \and J. C. Kasper\inst{3,9}
                                        \and R. J. MacDowall\inst{11}
                                        \and D. M. Malaspina\inst{12}
                                        \and M. Pulupa\inst{5}
                                        \and M. L. Stevens\inst{9}
                                        }

         \institute{LESIA, Observatoire de Paris, Université PSL, CNRS, 
Sorbonne Université, Université de Paris, 
5 place Jules Janssen, 92195 Meudon, France\\
                                                        \email{mingzhe.liu@obspm.fr}
                                 \and Department of Physics and Astronomy, University of Iowa, IA 52242, USA
                                 \and Climate and Space Sciences and Engineering, University of Michigan, Ann Arbor, MI 48109, USA
                                 \and IRAP, Universite Paul Sabatier, 9 Av du Colonel Roche, BP 4346, 31028, Toulouse cedex 4, France
                                 \and Space Sciences Laboratory, University of California, Berkeley, CA 94720-7450, USA
                                 \and Physics Department, University of California, Berkeley, CA 94720-7300, USA
                                 \and The Blackett Laboratory, Imperial College London, London, SW7 2AZ, UK
                                 \and School of Physics and Astronomy, Queen Mary University of London, London E1 4NS, UK
                                 \and Smithsonian Astrophysical Observatory, Cambridge, MA 02138 USA
                                 \and School of Physics and Astronomy, University of Minnesota, Minneapolis, MN 55455, USA
                                 \and Solar System Exploration Division, NASA/Goddard Space Flight Center, Greenbelt, MD, 20771 
                                 \and Laboratory for Atmospheric and Space Physics, University of Colorado, Boulder, CO 80303, USA                    
                                }

         \date{Received: October 8 2020; Accepted: December 29 2020}

        \abstract
         {} 
         {We investigate the solar wind energy flux in the inner heliosphere using 12-day observations around each perihelion of Encounter One (E01), Two (E02), Four (E04), and Five (E05) of Parker Solar Probe (PSP), respectively, with a minimum heliocentric distance of 27.8 solar radii ($R_\odot{}$).} 
         {Energy flux was calculated based on electron parameters (density $n_e$, core electron temperature $T_{c}$, and suprathermal electron temperature $T_{h}$) obtained from the simplified analysis of the plasma quasi-thermal noise (QTN) spectrum measured by RFS/FIELDS and the bulk proton parameters (bulk speed $V_p$ and temperature $T_p$) measured by the Faraday Cup onboard PSP, SPC/SWEAP.}
         {Combining observations from E01, E02, E04, and E05, the averaged energy flux value normalized to 1 $R_\odot{}$ plus the energy necessary to overcome the solar gravitation ($W_{R_{\sun}}$) is about 70$\pm$14 \wm{}, which is similar to the average value (79$\pm$18 \wm{}) derived by Le Chat et al from 24-year observations by Helios, Ulysses, and Wind at various distances and heliolatitudes. It is remarkable that the distributions of $W_{R_{\sun}}$ are nearly symmetrical and well fitted by Gaussians, much more so than at 1 AU, which may imply that the small heliocentric distance limits the interactions with transient plasma structures.}
         {}

          \keywords{(Sun:) solar wind---Sun: heliosphere---Sun: corona---Sun: fundamental parameters---plasmas---acceleration of particles}

         \maketitle

%

\section{Introduction} \label{1}

The question of how the solar wind is produced and accelerated is unsolved since its discovery about sixty years ago \citep{1958Parker, 1962Neugebauer} and \cite{2001Parker} showed that "we cannot state at the present time why the Sun is obliged by the basic laws of physics to produce the heliosphere". An important property of the solar wind is its energy flux, which is similar in the whole heliosphere and in the fast and slow wind \citep[e.g.,][]{1990Schwenn, 2006MeyerVernet, 2009LeChat, 2012LeChat}, and much more so than the particle flux itself. As shown by \cite{2009LeChat}, the energy flux is of a similar fraction of the luminosity for Solar-like and cool giant stars, which suggests that stellar winds may share a basic process for their origin and acceleration. Investigations of the solar wind energy flux in the inner heliosphere are of significant importance for astrophysics, but there are still very few of them.

\cite{2006MeyerVernet,2007meyervernet} showed that the average solar wind energy flux scaled to one solar radius of about 70 \wm{} from long-term Helios and Ulysses observations is close to the average total energy flux of solar flares and $10^{-6}$ times the solar luminosity -- a fraction similar to that of a number of other stars. With a much larger solar wind data set from several spacecraft at various distances and latitudes, \cite{2012LeChat} found an average value of 79$\pm$18 \wm{} between 1976 and 2012, whereas \cite{2014McComas} found an average value of about 60 \wm{} with OMNI data at 1 AU between 2011 and 2014. Helios 1 and 2 orbits ranged from 0.3 to 1 AU \citep{1975Schwenn}, whereas Ulysses operated between 1 and 4 AU \citep{1992Wenzel}. The ongoing, pioneering mission of Parker Solar Probe (PSP) \citep{2016Fox} orbits with perihelia of heliocentric distances decreasing from 35.7 solar radii ($R_\odot{}$) to 9.86 $R_\odot{}$ within five years. Four instruments onboard PSP, including the Fields experiment (FIELDS) \citep{Bale2016}, Solar Wind Electrons Alphas and Protons investigation (SWEAP) \citep{Kasper2016}, Integrated Science Investigation of the Sun (IS$\odot{}$IS) \citep{McComas2016}, and Wide-field Imager for Solar PRobe (WISPR) \citep{Vourlidas2016}, are working together to provide both in situ and remote observations. In situ field and plasma measurements of the inner heliosphere from FIELDS/PSP and SWEAP/PSP offer an opportunity to estimate the solar wind energy flux closer to the Sun than previously derived.

FIELDS/PSP provides accurate electron density and temperature measurements via quasi-thermal noise (QTN) spectroscopy. This technique has been used in a number of space missions \citep[e.g.,][]{1979meyervernet,1986meyervernet,2017meyervernet,1999Issautier,2001Issautier,2008Issautier,1995Maksimovic,2005Moncuquet,2006Moncuquet}, and it is an effective and efficient tool. Recently, \cite{2020Moncuquet} and \cite{Maksimovic2020} derived preliminary solar wind electron measurements from the plasma QTN spectra observed by the Radio Frequency Spectrometer (RFS/FIELDS) \citep[see][]{2017Pulupa}. SWEAP/PSP consists of the Solar Probe Cup (SPC) and the Solar Probe Analyzers (SPAN) \citep{Kasper2016, Case2020, Whittlesey_2020}. SPC is a fast Faraday cup designed to measure the one dimensional velocity distribution function (VDF) of ions and sometimes electrons and SPAN is a combination of three electrostatic analyzers operated to measure the three dimensional VDFs of ions and electrons. Due to the instrument design, the SPAN-Ai instrument cannot observe the complete core of the solar wind ions in the first several encounters and SPC can provide ion observations during SPAN's observational gaps by pointing at the Sun during the encounter phase of each orbit, although SPC sometimes cannot detect the whole distribution \citep{Kasper2016, Whittlesey_2020, Case2020}.

Therefore, we calculated the solar wind energy flux with both the RFS/FIELDS/PSP (electron) and SPC/SWEAP/PSP (ion) observations during Encounters One (E01), Two (E02), Four (E04), and Five (E05) (Section \ref{2}). The minimum heliocentric distance is 35.66 $R_\odot{}$ for E01 and E02 and around 27.8 $R_\odot{}$ for E04 and E05. In Section \ref{3}, we analyze the relationship between the energy flux, the bulk speed, and the plasma density (Section \ref{3.1}). How the total energy flux and each component of it evolve with increasing heliocentric distance is studied in Section \ref{3.2}. In Section \ref{4}, the results are summarized and discussed.

\section{Data analysis} \label{2}

The solar wind energy flux ($W$), which includes the kinetic energy ($W_{kinetic}$), the enthalpy ($W_{enthalpy}$), and the heat flux ($Q$), is expressed as 

\begin{equation}\label{e1}
W = W_{kinetic} + W_{enthalpy} + W_g + Q
\end{equation}

where we have neglected the wave energy flux and added the flux equivalent to the energy required to overcome the solar gravitation $W_g$ \citep{1990Schwenn}; Q is the sum of the electron heat flux $q_e$ and proton heat flux $q_p$. \cite{2020Halekasb} and \cite{2020Halekasa} found that $q_e$ ranges from $10^{-4}$ to $10^{-3}$ \wm{} during E01, E02, E04, and E05 of PSP orbits, which can be neglected (See section \ref{3}). We note that at 1 AU, $q_e$ measured with Helios is $q_e\approx10^{-6}$ \wm{} \citep{1990Pilipp}, while $q_p$ ranges from about $10^{-7}$ (1 AU) to $10^{-5}$ (0.3 AU) \wm{} \citep{2011Hellinger}. We therefore neglected both the electron and proton heat flux compared to the other components, so that

\begin{equation}\label{e2}
W = W_{kinetic} + W_{enthalpy} + W_g
\end{equation}

where the expressions of the different components are given below. It is important to note that \cite{2012LeChat} neglected the enthalpy at 1 AU. However, this contribution cannot be ignored closer to the Sun, where it contributes to about 5$\%$ of the total energy flux (See section \ref{3.2}): 

\begin{equation}\label{e4}
\begin{aligned}
W_{kinetic} & = n_pm_pV_p\frac{{V_p}^2}{2}+ n_{\alpha}m_{\alpha}V_{\alpha}\frac{{V_{\alpha}}^2}{2}
\end{aligned}
\end{equation}

\begin{equation}\label{e5}
\begin{aligned}
W_{enthalpy} & = n_eV_p\frac{5k_B T_e}{2} + n_pV_p\frac{5k_B T_p}{2} +n_{\alpha}V_{\alpha}\frac{5k_B T_{\alpha}}{2}\\
                                                          & \approx n_eV_p\frac{5k_B T_e}{2} + n_pV_p\frac{5k_B T_p}{2}
\end{aligned}
\end{equation}

\begin{equation}\label{e6}
\begin{aligned}
W_g & = (n_pm_pV_p\frac{GM_{\sun}}{R_{\sun}} + n_{\alpha}m_{\alpha}V_{\alpha}\frac{GM_{\sun}}{R_{\sun}})(1-\frac{R_{\sun}}{r})
\end{aligned}
.\end{equation}

Here, $n_p$, $m_p$, $n_{\alpha}$, and $m_{\alpha}$ denote the proton number density, proton mass, $\alpha$ particle number density, and $\alpha$ particle mass, respectively. Furthermore, $V_{p}$ ($V_{\alpha}$) is the solar wind proton ($\alpha$) bulk speed, $n_e$ is the electron number density, $k_B$ is the Boltzmann constant, $T_p$ ($T_e$) is the proton (electron) temperature, $G$ is the gravitational constant, $M_{\sun}$ is the solar mass, $R_{\sun}$ is the solar radius, and $r$ is the heliocentric distance of PSP. We note that $T_e$ was derived from the core electron temperature $T_{c}$ and suprathermal electron temperature $T_{h}$ with $T_e = T_c + (n_h/n_e)T_h$, where $n_h$ denotes the suprathermal electron density and $n_h/n_e$ is assumed to be 0.1 \citep[see][]{2020Moncuquet, 2009JGRA}. In Equation \ref{e4}, \ref{e5}, and \ref{e6}, we assume that $V_{\alpha} \approx V_{p}$ and ignore the enthalpy of the $\alpha$ particles since $n_{\alpha}$ is much smaller than $n_e$ (and both $V_{\alpha}$ and $T_{\alpha}$ are not available). The energy flux was scaled to one solar radius as written below, yielding the total energy required at the base to produce the wind -- a basic quantity for understanding the wind production and comparing the Sun to other wind-producing stars:  

\begin{equation}\label{e3}
W_{R_{\sun}} = W(r)\frac{r^2}{{R_{\sun}}^2}
.\end{equation}

We used the level-3 ion data (moments) from SPC/SWEAP \citep{Kasper2016, Case2020} and the electron parameters deduced from the simplified QTN method with the observations from RFS/FIELDS \citep{2020Moncuquet, 2017Pulupa}. For each encounter, only 12-day high-time-resolution observations near the perihelion were considered: SPC collects one sample or more every 0.874 seconds and the QTN datasets have a 7-sec resolution. Since the resolution of the datasets from SPC is different from that of the QTN datasets, we interpolated them to the same resolution to carry out the calculations. Currently, $\alpha$ particle observations directly obtained from SPC/SWEAP cannot be used due to calibration issues. Also, $n_p$ is too different from $n_e$ (being smaller than $n_e$ by more than 30\% on average) with an estimation of $<$$n_{\alpha}/n_e$$>$$=$$<$$(n_e-n_p)/(2 \times n_e)$$>$ $\approx 16.0 \%$, which implies unrealistic values for $n_{\alpha}$ obtained based on plasma neutrality. Past studies \citep[e.g.,][]{2007Kasper, 2012Kasper, 2019Alterman, 2020Alterman} show that the $\alpha$ particle abundance ($A_{He}= n_{\alpha}/n_p$) rarely exceeds $A_{He}$$\sim$5\%, especially when the bulk speed of the solar wind is below $V_p = 400$ \kms{}. \cite{2020Alterman} show that at 1 AU, $A_{He}$ ranges from 1\% to 5\% during Solar Cycle 23 and 24 and predict that 1\% $<$ $A_{He}$ $<$ 4\% at the onset of Solar Cycle 25 (solar minimum). We assume that $A_{He}$ (which is almost the same as $n_{\alpha}/n_e$) of the solar wind remains the same when it propagates from the inner heliosphere to 1 AU \citep{2020Viall}. As a result, we deduced $n_{\alpha}$ with $n_e$ where $n_{\alpha}/n_e$ is a free parameter ranging from 1\% to 4\% \citep{2020Alterman}. This enabled us to determine $n_p$ based on the plasma neutrality. The resulting values of $n_{\alpha}$ and $n_p$ were used to calculate $W$ and then $W_{R_{\sun}}$.

\section{Observations and results} \label{3}

During the first and second encounter of PSP, it reached the perihelion of 35.66 $R_\odot{}$ ($\sim$ 0.17 AU) on November 6, 2018 and April 5, 2019, respectively. For both E04 and E05, PSP arrived at the perihelion of 27.8 $R_\odot{}$ ($\sim$ 0.13 AU) on January 29, 2020 and June 7, 2020, respectively. In Section \ref{3.1}, we give an overview of the PSP measurements of solar wind density, speed, and energy flux for all available encounters including E01, E02, E04, and E05. We note that E03 observations are not considered due to the lack of SPC observations near the perihelion. For each encounter, 12-day observations around the perihelion were used for calculations. The heliocentric distance for both E01 and E02 ranges from 35.66 to about 55 $R_\odot{}$, and it ranges from 27.8 to about 57 $R_\odot{}$ for both E04 and E05. In Section \ref{3.2}, we combine the observations from E01, E02, E04, and E05 to show the histogram distributions and the evolution of the energy flux as a function of heliocentric distance.  

\subsection{Overview of E01, E02, E04, and E05} \label{3.1}

\begin{figure*}
\centering
                \includegraphics[width=0.65\textwidth]{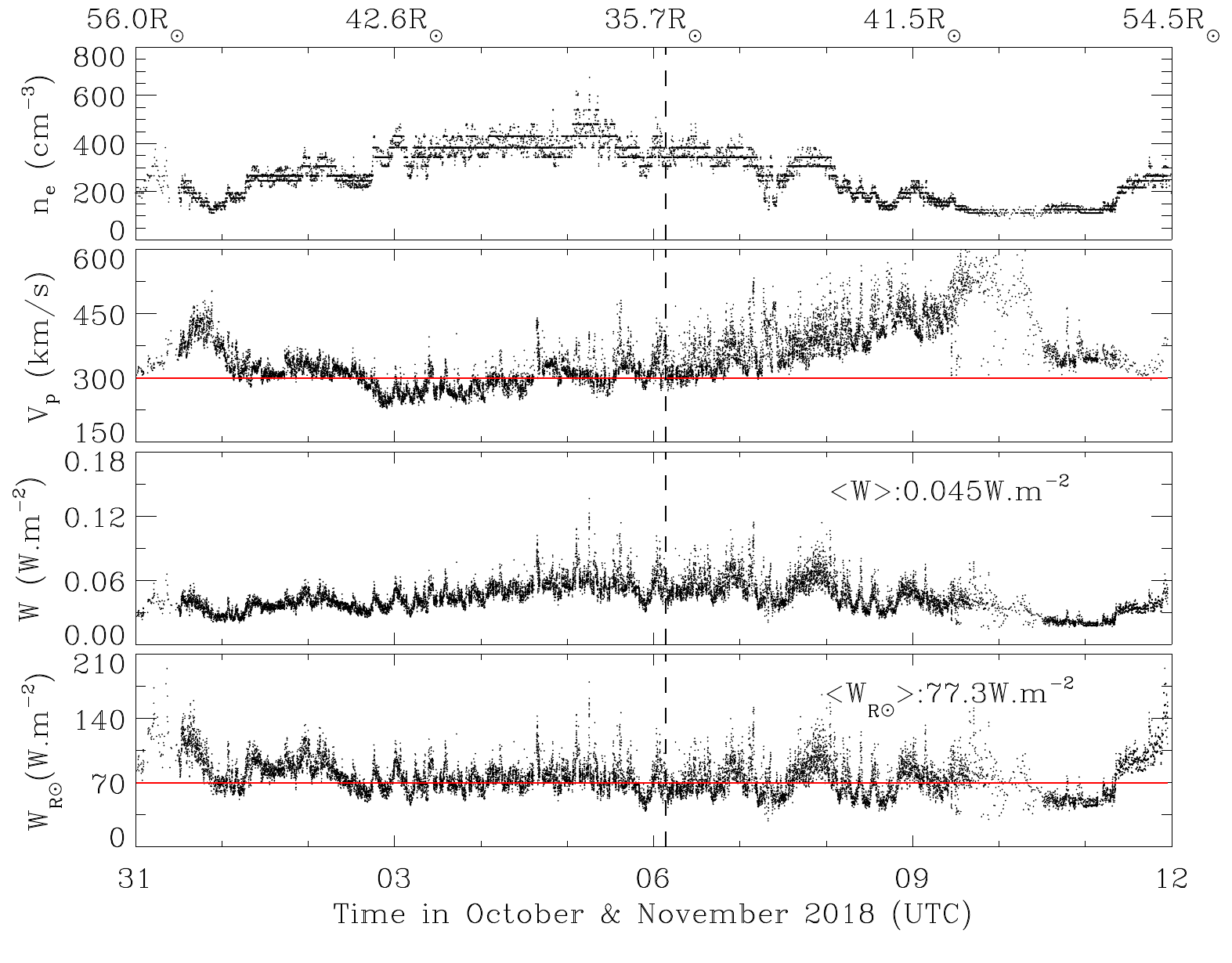}
                \caption{Solar wind density, speed, and energy flux measurements by PSP during Encounter One (from October 31, 2018 00:00:00 to November 12, 2018 00:00:00 UTC). First panel: QTN electron density. Second panel: Proton bulk speed. A red horizontal line ($V_p= 300$ \kms{}) is plotted for reference. Third panel: Solar wind energy flux $W$. Fourth panel: Solar wind energy flux normalized to one solar radius $W_{R_{\sun}}$ (black) with a red horizontal line ($W_{R_{\sun}}= 70$ \wm{}) superimposed for reference. The heliocentric distance (in units of the Solar radius $R_{\sun}$ ) is indicated at the top of the first panel and the black vertical line denotes the perihelion of the PSP orbit.}
                \label{f1}%
\end{figure*}

Figure~\ref{f1} shows an overview of the PSP measurements of solar wind density, speed, and energy flux during E01 (from October 31, 2018 00:00:00 to November 12, 2018 00:00:00 UTC). The top panel presents the electron number density ($n_e$) obtained by the QTN method. In the second panel, the proton bulk speed is shown. The third and fourth panels present the solar wind energy flux ($W$, from equation \ref{e2}) and its value scaled to one solar radius ($W_{R_{\sun}}$, from equation \ref{e3}), respectively. In Figure~\ref{f1}, $n_{\alpha}$ and $n_p$ were computed from $n_e$ based on $n_{\alpha}/n_e = $ 2.5$\%$ for calculating $W$ and $W_{R_{\sun}}$. Most of the time, $V_p$ varies around 300 \kms{}, and $W_{R_{\sun}}$ varies around 70 \wm{}. The average values of $W$ and $W_{R_{\sun}}$ are  0.045 \wm{} and 77.3 \wm{}, respectively. The average value of $W_{R_{\sun}}$ of E01 is consistent with the long-term observations from \cite{2012LeChat} (around 79 \wm{}). We note that $W_{R_{\sun}}$ does not vary much with $V_p$ when $V_p$ increases abruptly (i.e., from November 8 to 10, 2018).

\begin{figure*}
\centering
                \includegraphics[width=0.65\textwidth]{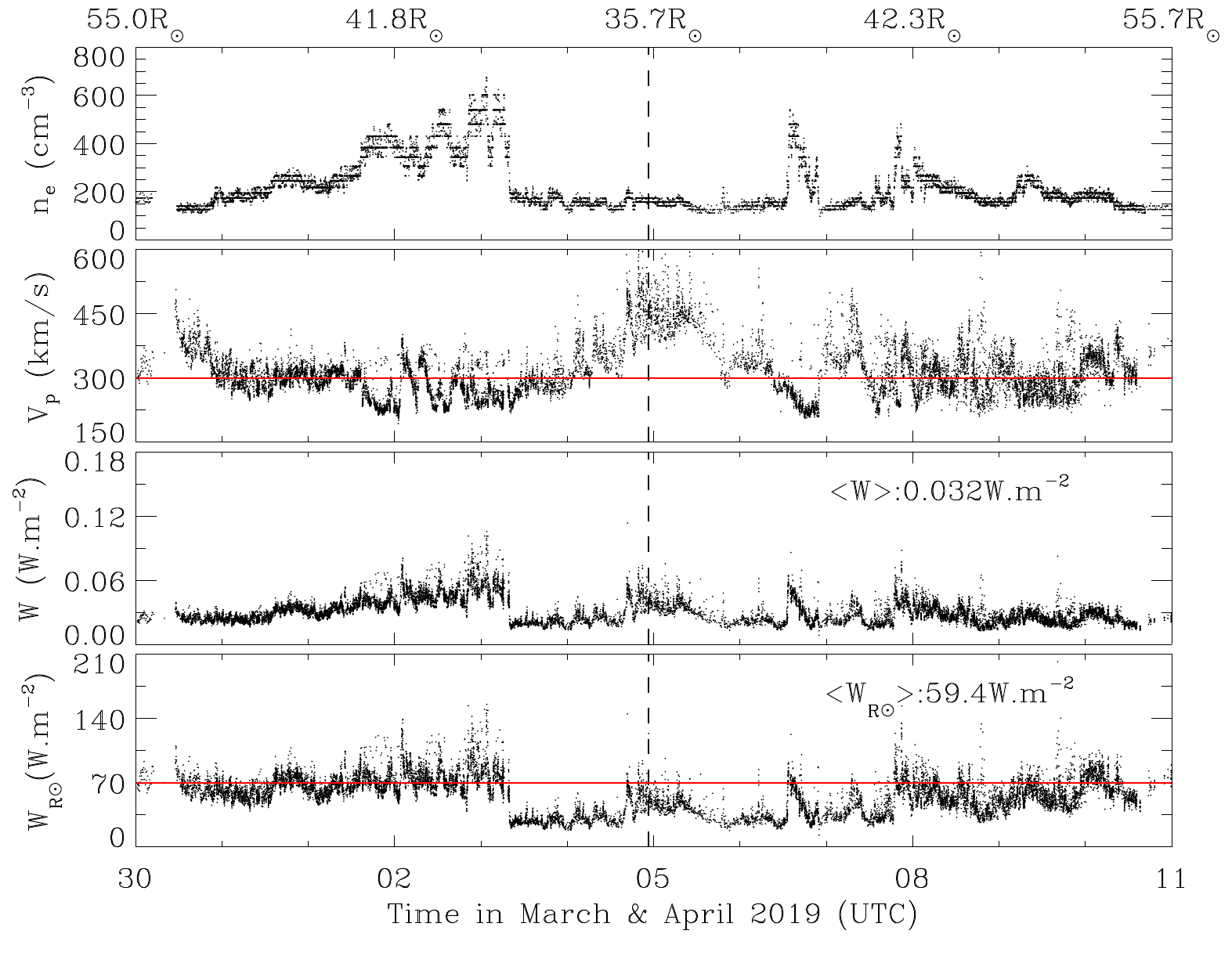}
                \caption{Solar wind density, speed, and energy flux measurements by PSP for Encounter Two (March 30, 2019 00:00:00 to April 11, 2019 00:00:00 UTC). This figure follows the same format as that of Figure~\ref{f1}.} 
                \label{f2}%
\end{figure*}

Figure~\ref{f2}, which follows the same format as Figure~\ref{f1}, summarizes the PSP measurements of solar wind density, speed, and energy flux during E02 (from March 30, 2019 00:00:00 to April 11, 2019 00:00:00 UTC).  We deduced $n_p$ and $n_{\alpha}$ with the same method used for E01 to calculate both $W$ and $W_{R_{\sun}}$. We note that $n_e$ shows two successive low plateaus near the perihelion of E02 (from April 3 to 8, 2019 UT), as shown in the first panel of Figure~\ref{f2}, whereas $V_p$ shows two high peaks. This is in agreement with the well-known anticorrelation between the solar wind speed and density \citep[e.g.,][]{1996Richardson,2012LeChat}. Both $W_{R_{\sun}}$ and $W$ also show two low plateaus near the perihelion of E02 (from April 3 to 8, 2019 UT), similar to the solar wind density. Elsewhere, $V_p$ remains around 300 \kms{} and $W_{R_{\sun}}$ varies around 70 \wm{}. The mean values of $W$ and $W_{R_{\sun}}$ during E02 are 0.032 \wm{} and 59.4 \wm{}, respectively.

\begin{figure*}
\centering
                \includegraphics[width=0.65\textwidth]{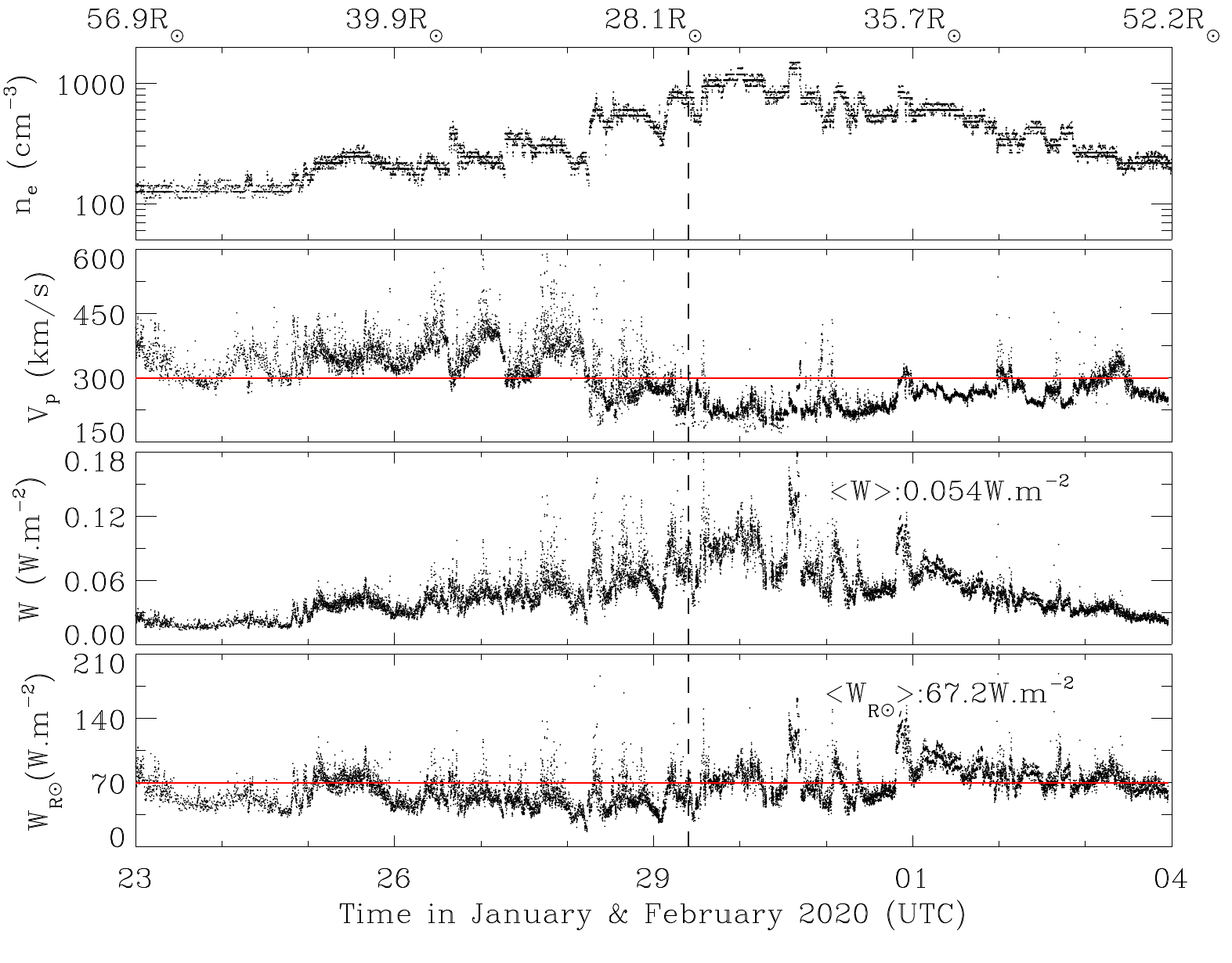}
                \caption{Solar wind density, speed, and energy flux measurements by PSP for Encounter Four (from January  23, 2020 00:00:00 to February 4, 2020 00:00:00 UTC), which follows the same format as that of Figure~\ref{f1}.} 
                \label{f3}%
\end{figure*}

Similarly, Figure~\ref{f3} illustrates the PSP observations during E04 (from January 23, 2020 00:00:00 to February 4, 2020 00:00:00 UTC). We used $n_p$ and $n_{\alpha}$, which were deduced with the same method used for both E01 and E02, when calculating both $W$ and $W_{R_{\sun}}$. The second panel of Figure~\ref{f3} shows that $V_p$ varies around 375 \kms{} before January 29, 2020 and is predominantly 225 \kms{} afterward. Furthermore, $W_{R_{\sun}}$ varies around 70 \wm{} and does not change significantly even when $V_p$ decreases sharply from January 28 to 30, 2020. The average values of $W$ and $W_{R_{\sun}}$ for E04 are 0.054 \wm{} and 67.2 \wm{}, respectively.

\begin{figure*}
\centering
                \includegraphics[width=0.65\textwidth]{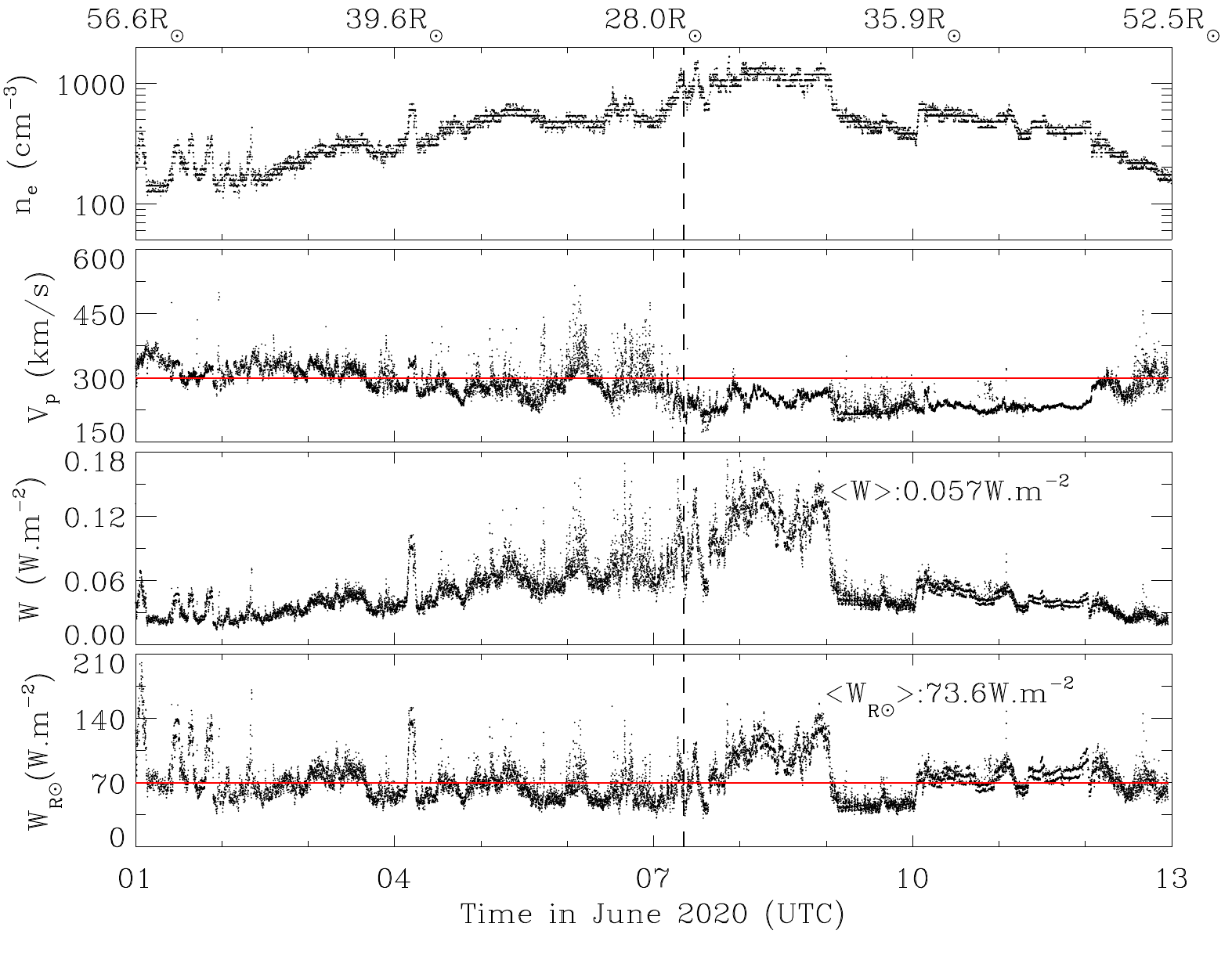}
                \caption{Solar wind density, speed, and energy flux measurements by PSP for Encounter Five (from June 1, 2020 00:00:00 to June 13, 2020 00:00:00 UTC), which follows the same format as that of Figure~\ref{f1}.} 
                \label{f4}
\end{figure*}

Figure~\ref{f4} is similar to Figure~\ref{f1},~\ref{f2}, and~\ref{f3}, but for E05 (from June 1, 2020 00:00:00 to June 13, 2020 00:00:00 UTC). We used the same method as previously explained for E01, E02, and E04 for calculating the energy flux. During this encounter, $V_p$ usually stays at around 300 \kms{} except from June 7 to 12, 2020 during which $V_p$ remains approximately at 225 \kms{}. For E05, $W_{R_{\sun}}$ is predominantly about $W_{R_{\sun}}=70$ \wm{}. From June 7 to 10, 2020, both $W$ and $W_{R_{\sun}}$ experience sharp changes, which results from a sharp variation in $n_e$. The corresponding values of both $W$ and $W_{R_{\sun}}$ are larger (smaller) than the ambient values at the beginning (in the end) of this time period. The average values of $W$ and $W_{R_{\sun}}$ for E05 are 0.057 \wm{} and 73.6 \wm{}, respectively.


\begin{table}
\caption{Energy flux average value of each encounter.}             
\label{table:1}      
\centering                          
\begin{tabular}{c c c c c}        
\hline\hline                 
Energy Flux (\wm)     & E01   & E02   & E04   & E05   \\    
\hline                        
$<$$W$$>$             & 0.045 & 0.032 & 0.054 & 0.057 \\
$<$$W_{R_{\sun}}$$>$  & 77.3  & 59.4  & 67.2  & 73.6  \\  
\hline                                   
\end{tabular}
\end{table}

 Table \ref{table:1} summarizes the average values of the energy flux $<$$W$$>$ and the values normalized to one solar radius $<$$W_{R_{\sun}}$$>$ for the four PSP encounters mentioned above. We note that the sequence difference between $<$$W_{R_{\sun}}$$>$ and $<$$W$$>$ results from the $r^{-2}$ normalization when deriving $W_{R_{\sun}}$, whereas the individual flux tubes vary differently. It is remarkable that these values of $<$$W_{R_{\sun}}$$>$ are close to those found previously \citep{2006MeyerVernet,2012LeChat} despite the smaller time durations and latitude extensions of PSP observations. We note the relatively low $<$$W_{R_{\sun}}$$>$ of E02 and the low solar wind density near the perihelion of PSP orbit (see Figure~\ref{f2}). The dilute transient solar wind structure observed around the perihelion helps to explain this relatively low value compared to the long-term observations of \cite{2012LeChat}. The origins of the low plateaus of plasma density related to high peaks of bulk speed are discussed by \cite{2020Rouillard} and they are outside the scope of this paper. \cite{2012LeChat} averaged the values over a solar rotation ($\sim$27.2 days) to reduce the effect of transient events such as coronal mass ejections (CMEs) or corotating interaction regions (CIRs). Although CMEs or small-scale flux ropes are observed by PSP during E01 \citep[e.g.,][]{2020Hess, 2020Zhao, 2020Korreck}, $<$$W_{R_{\sun}}$$>$ of E01 (77.3 \wm{}) is almost the same as the long-term averaged value found by \cite{2012LeChat}.

\subsection{Distributions of energy flux and variation with distance} \label{3.2}


\begin{figure*}
                \centering
                \begin{tabular}{cl}
                \includegraphics[width=0.33\textwidth]{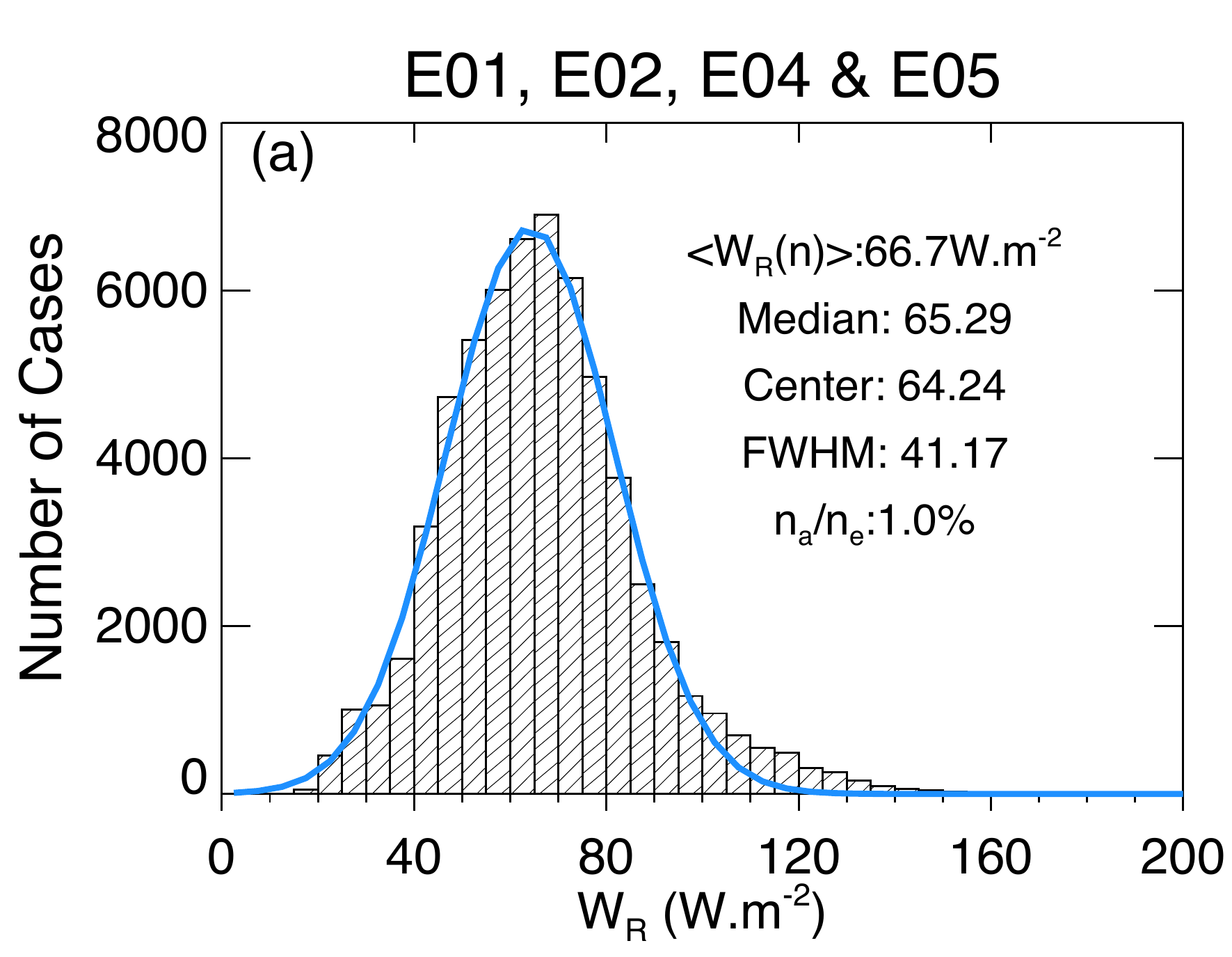}
                \includegraphics[width=0.33\textwidth]{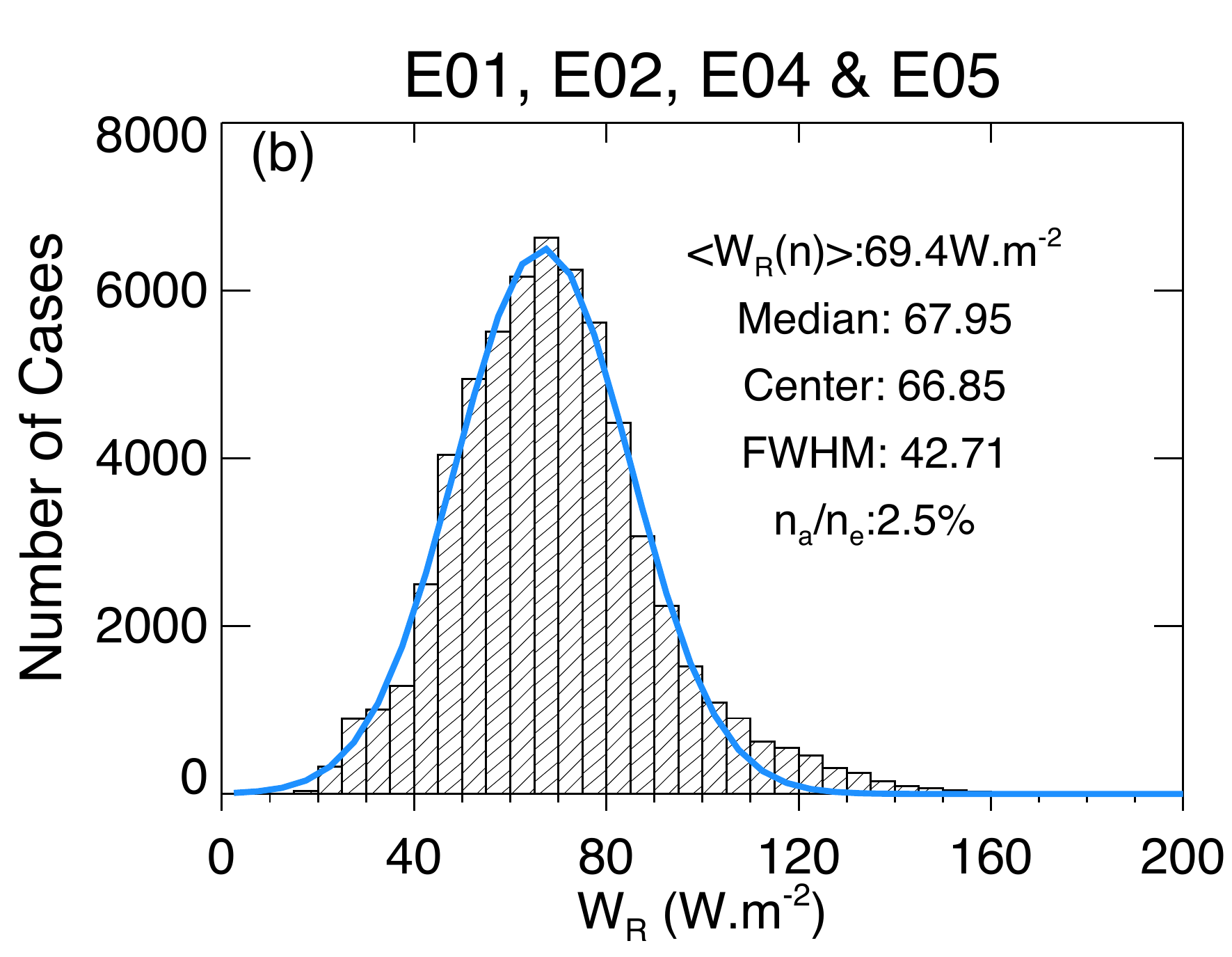}
                \includegraphics[width=0.33\textwidth]{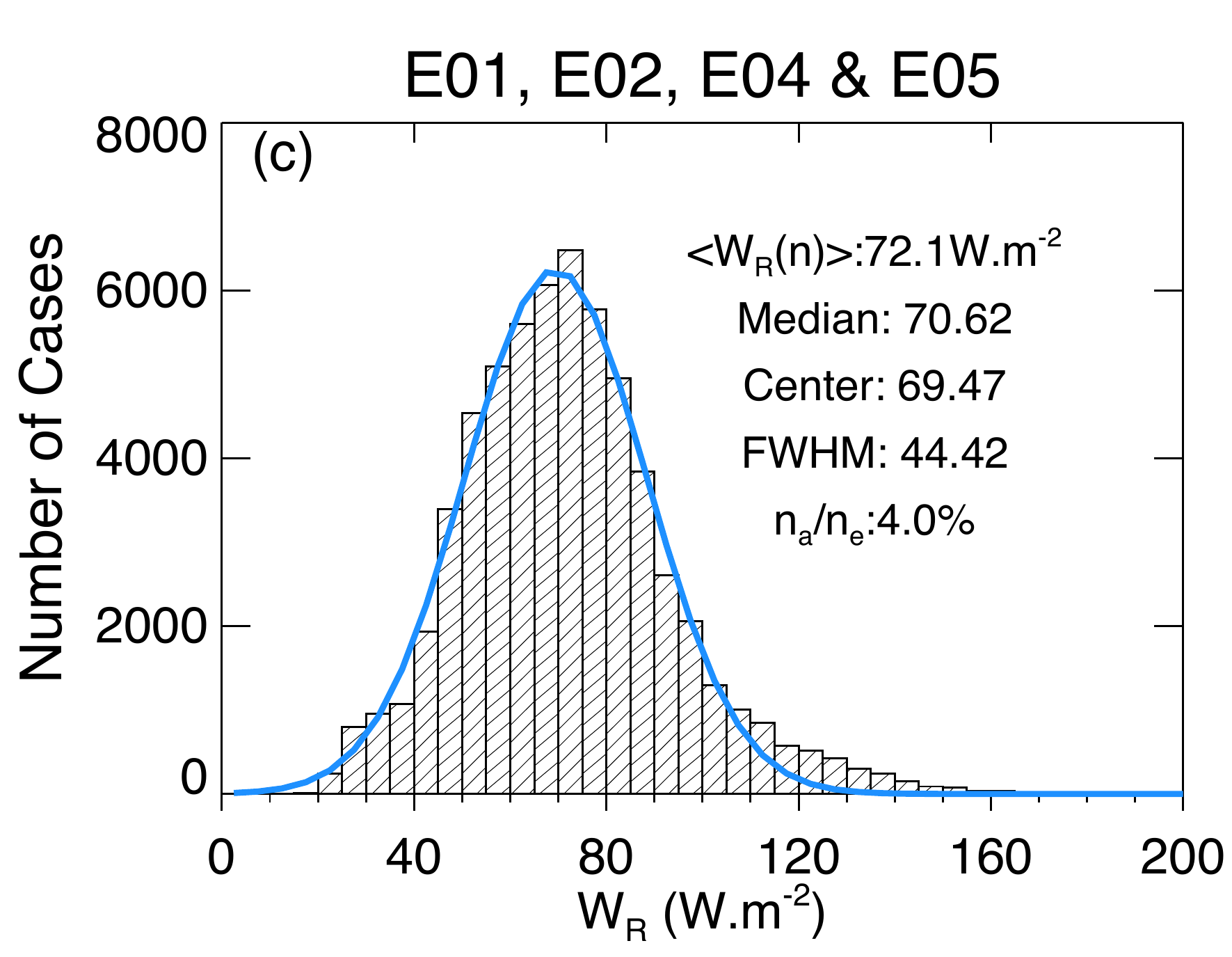}\\
                \end{tabular}
                \caption{Distributions of solar wind energy flux ($W_R$) normalized to one solar radius with a ratio between $\alpha$ particle number density ($n_{\alpha}$) and electron number density ($n_e$) ranging from 1$\%$ to 4$\%$ for Encounters E01, E02, E04, and E05. Figs (a), (b), and (c) assume $n_{\alpha}/n_e=$ 1$\%$, 2.5$\%$, and 4$\%$, respectively, to illustrate the uncertainty due to the absence of $\alpha$ measurements. Average and median values of each histogram are indicated with Gaussian fits superimposed in blue. Center value and standard deviation (full-width-half-maximum) of the Gaussian fit are also presented.}
                \label{f5}
\end{figure*}

Figure~\ref{f5} shows the distributions of $W_{R_{\sun}}$ combining the observations from E01, E02, E04, and E05. Based on the assumption that $n_{\alpha}/n_e$ ranges from 1.0$\%$ to 4.0$\%$, we calculated $W_{R_{\sun}}$ with $n_{\alpha}/n_e = $ 1.0$\%$, 2.5$\%$, and 4.0$\%$ and the corresponding results are shown in Figure~\ref{f5} (a), (b), and (c), respectively. Each histogram distribution was fitted with a Gaussian function (blue line), and the center value (the most probable value) and standard deviation (full-width-half-maximum which is short for FWHM) are shown together with the mean and median values. It is remarkable that the histograms of $W_{R_{\sun}}$ are very symmetrical and nearly Gaussian. The difference between the average, median, and most probable fit value of $W_{R_{\sun}}$ is very small (less than 3$\%$). With a fixed $n_{\alpha}/n_e$ ratio, the uncertainties of $<$$W_{R_{\sun}}$$>$ resulting from the uncertainties of the plasma parameters $n_e$, $V_p$, $T_e$, and $T_p$ are 10.0$\%$, 4.1$\%$, 0.85$\%$, and 0.28$\%$, respectively. We used the uncertainty of $n_e$ provided by the QTN method and \cite{2020Moncuquet} estimate that the uncertainty of $T_e$ is around 20$\%$. \cite{Case2020} share that the estimated uncertainties of $V_p$ and $T_p$ are 3.0$\%$ and 19$\%$, respectively. When $n_{\alpha}/n_e$ increases from 1.0$\%$ to 2.5$\%$ and then to 4.0$\%$, $<$$W_{R_{\sun}}$$>$ increases from 66.7 \wm{} to 69.4 \wm{} and then to 72.1 \wm{}, and the values of FWHM increase from 41.2 \wm{} to 42.7 \wm{} and then to 44.4 \wm{}. The uncertainty of $W_{R_{\sun}}$ resulting from the variation of $n_{\alpha}/n_e$ is around 4\%. Furthermore,  $<$$W_{R_{\sun}}$$>$ from the E01, E02, E04, and E05 observations is around 69.4 \wm{} with a total uncertainty that we estimate to be at most 20.0\%, which is consistent with previous results \citep[e.g.,][]{1990Schwenn,2006MeyerVernet,2009LeChat,2012LeChat,2014McComas}.

\begin{figure*}
\centering
                \includegraphics[width=0.5\textwidth]{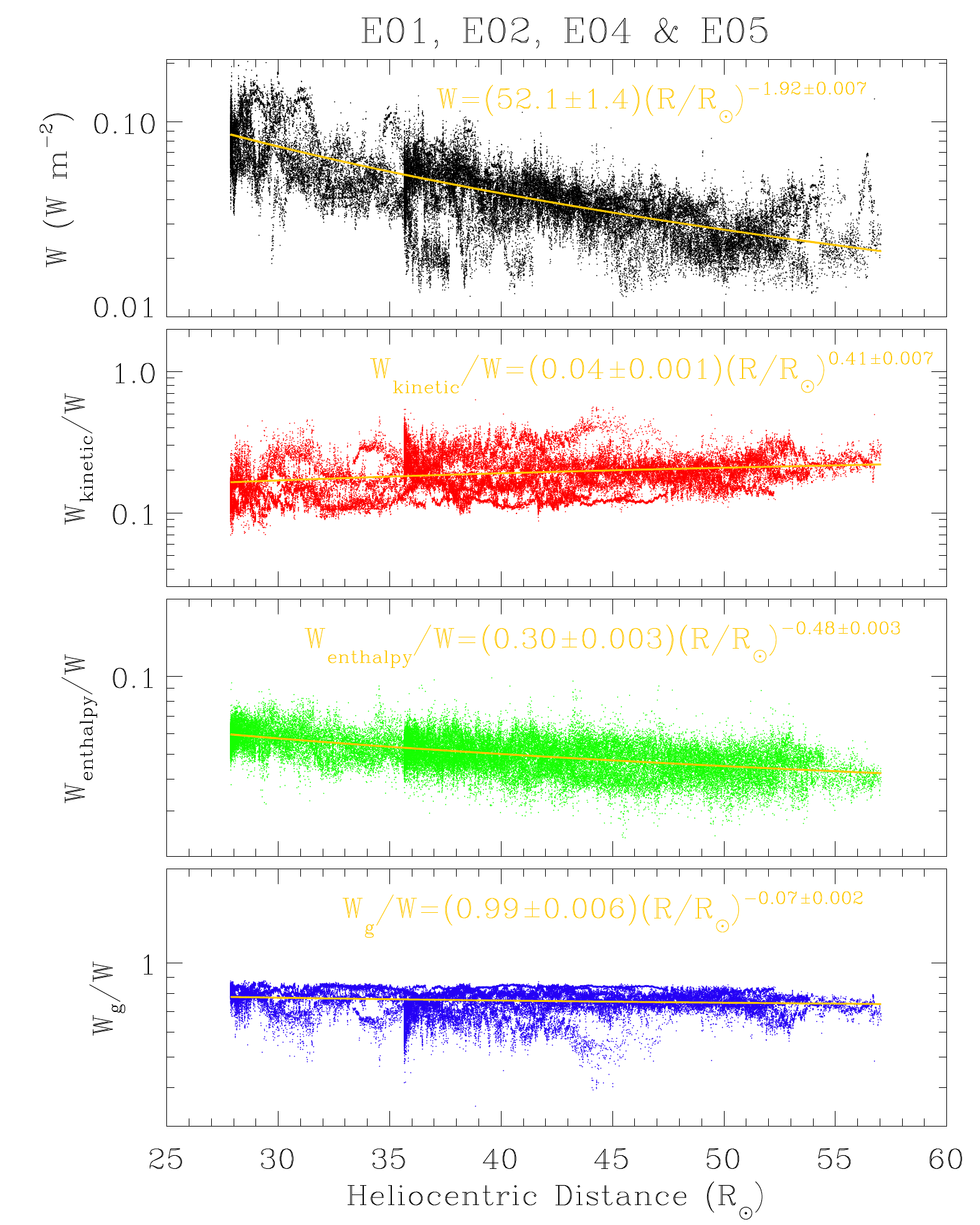}
                \caption{Variation of $W$ and its components with heliocentric distance combining observations from Encounter One (E01), Two (E02), Four (E04), and Five (E05). From top to bottom, evolution of $W$, $W_{kinetic}/W$, $W_{enthalpy}/W$, and $W_g/W$ with heliocentric distance are shown, respectively. The fitted profile (yellow) is superimposed on each corresponding panel, respectively.} 
                \label{f6}
\end{figure*}

Figure~\ref{f6} presents $W$, $W_{kinetic}/W$, $W_{enthalpy}/W$, and $W_g/W$ as a function of heliocentric distance in units of solar radius $R_{\sun}$, which includes the observations from E01, E02, E04, and E05. Levenberg-Marquardt least-squares fit was performed on each quantity and the fitted functions are shown in the figure. We note that the power index for $W$ is -1.92 (near to -2.0), which is in agreement with Equation \ref{e3} used to scale the solar wind energy flux to one $R_{\sun}$. When PSP moves from 57.1 $R_\odot{}$ to 27.8 $R_\odot{}$, $W_{kinetic}$, in order of magnitude, ranges from $10^{-3}$ to $10^{-2}$ \wm{}, while $W_{enthalpy}$ and $W_g$ range from $10^{-3}$ to $10^{-2}$ \wm{} and from $10^{-2}$ to $10^{-1}$ \wm{}, respectively. Further, as shown in Figure~\ref{f6}, $W_g$ is the dominant term for $W$, $W_{kinetic}$ is the second most dominant one, and $W_{enthalpy}$ is the least dominant term. Even though the contribution of $W_{enthalpy}$ to $W$ is still the least among the three components in the inner heliosphere, it reaches about 30\% of the kinetic energy flux at the smallest distances and we cannot neglect it directly ($<$$W_{enthalpy}$$>$$/$$<$$W$$>$$\approx$ 5$\%$). We note that since $W_g$ exceeds $W_{kinetic}$ by a factor of about four, most of the energy supplied by the Sun to generate the solar wind serves to overcome the solar gravity. As is shown in the first panel of Figure~\ref{f6}, the energy flux can reach $W \approx 10^{-1}$ \wm{} near the perihelia of PSP orbits, whereas the corresponding electron heat flux is $q_e \approx 10^{-3}$ \wm{} \citep[see][]{2020Halekasb,2020Halekasa}. At most, $q_e$ contributes to 1.0$\%$ of $W$, and proton heat flux $q_p$ is usually much less than $q_e$. Therefore, neglecting the heat flux does not affect the conclusions made in this work.

\section{Discussion and conclusions} \label{4}

This paper presents the first analysis of the solar wind energy flux in the inner heliosphere (adding the flux equivalent to the energy necessary to move the wind out of the solar gravitational potential) with PSP observations. This covers heliocentric distances from 0.13 AU ($\sim$27.8 $R_{\sun}$) to 0.27 AU ($\sim$57.1 $R_{\sun}$) in combination of data during E01, E02, E04, and E05. This enables us to study the solar wind energy flux in the inner heliosphere, which is of great importance to understand the acceleration of the solar wind. We note that E03 is excluded due to the lack of SPC observations near perihelion. 

We find that the average value of $W_{R_{\sun}}$, $<$$W_{R_{\sun}}$$>$, is about 69.4 \wm{} with a total uncertainty of at most 20$\%$, which is similar to previous results based on long-term observations at greater distances and various latitudes \citep[e.g.,][]{1990Schwenn, 2006MeyerVernet, 2009LeChat, 2012LeChat,2014McComas}. This result confirms that this quantity appears as a global solar constant, which is of importance since it  is often used to deduce the solar wind density from the speed (or the reverse) in global heliospheric studies and modeling \citep[e.g.,][]{2018Shen,2014McComas,2017McComas,2020McComas,2019Krimigis,2020Wang}. 

It is remarkable that the distributions of $W_{R_{\sun}}$ are nearly symmetrical and well fitted by Gaussians. This may be explained by the limited interactions between solar wind and transient structures (e.g., CMEs and CIRs) in the inner heliosphere (below 0.27 AU).

Normalizing the solar wind energy flux as $1/r^{2}$ assumes a radial expansion of solar wind, which does not hold true for individual flux tubes, especially close to the Sun. However, this normalization holds true when integrating over a whole sphere surrounding the Sun, so that a large data set is necessary to obtain a reliable result. It is thus noteworthy that with only 12-day observations for each encounter (E01, E02, E04, and E05) and a limited latitude exploration, we find the same normalized energy flux as previous long-term studies at various latitudes. This is consistent with the fact that our dataset yields an energy flux varying with heliocentric distance with a power index close to -2. It is also interesting that this normalized energy flux represents a similar fraction of solar luminosity as observed for a large quantity of stars \citep{2006MeyerVernet,2012LeChat}. Since this quantity represents the energy flux to be supplied by the Sun for producing the wind \citep[e.g.,][]{1999MeyerVernet,2003Schwadron}, this similarity may provide clues to the physical processes at the origin of stellar winds \citep[e.g.,][]{2015Johnstonea}.

In this work, the heat flux was neglected when calculating the energy flux. When PSP gets much closer to the Sun, the contribution of the electron heat flux is larger \citep[see][]{2020Halekasb,2020Halekasa}. Furthermore, the solar wind protons often consist of two populations, that is to say core and beam drifting with respect to each other. The speed difference between them is typically on the order of the local Alfv{\'e}n speed \citep{2018Alterman}. It is likely that the proton heat flux will also be more important closer to the Sun. Therefore, the heat flux will be considered in a future work. Due to the lack of alpha particle observations, we make an assumption that $V_{\alpha} \approx V_{p}$. In fact, the differential speed between protons and alpha particles is also typically on the order of the local Alfv{\'e}n speed \citep[e.g.,][]{1996Steinberg,2017Durovcova,2018Alterman}, so that it may affect the energy flux closer to the Sun. We await more data that are to come in the future PSP encounters, with the recovery of the well calibrated alpha parameters.

\begin{acknowledgements}
                        The research was supported by the CNES and DIM ACAV+ PhD funding. Parker Solar Probe was designed, built, and is now operated by the Johns Hopkins Applied Physics Laboratory as part of NASA’s Living with a Star (LWS) program (contract NNN06AA01C). Support from the LWS management and technical team has played a critical role in the success of the Parker Solar Probe mission. We acknowledge the use of data from FIELDS/PSP (http://research.ssl.berkeley.edu/data/psp/data/sci/fields/l2/) and SWEAP/PSP (http://sweap.cfa.harvard.edu/pub/data/sci/sweap/).
\end{acknowledgements}

%
%

\bibliography{bibliography}
\bibliographystyle{aa} 

\end{document}